\documentstyle[aps,epsf]{revtex}		
\begin{document}
\twocolumn			
%
%
%
% VARIOUS DEFINITIONS:
%---------------------
%
\def\sigmap{\sigma^\prime}
\def\taup{\tau^\prime}
\def\sigmaconf{{\sigma_1,\ldots,\sigma_L}}
\def\tauconf{{\tau_1,\ldots,\tau_L}}
\def\sigmapconf{{\sigmap_1,\ldots,\sigmap_L}}
\def\taupconf{{\taup_1,\ldots,\taup_L}}
\def\doublesigconf{{\sigmaconf\,;\,\sigmapconf}}
\def\doubletauconf{{\tauconf\,;\,\taupconf}}
\def\transfermat{\mbox{\bf T}\,}
\def\sigstate{{\sigma_{j-m},\ldots,\sigma_{j+m}}}
\def\sigpstate{{\sigmap_{j-m},\ldots,\sigmap_{j+m}}}
\def\sigdouble{{\sigstate\,;\,\sigpstate}}
\def\sigidentical{{\sigstate\,;\,\sigstate}}
\def\taudouble{{\tau_j\,;\,\taup_j}}
\def\statenumber{s}
\def\selfunct{\chi}
\def\triplett{{\sigma_{i-1},\sigma_i,\sigma_{i+1}}}
\def\triplettp{{\sigmap_{i-1},\sigmap_i,\sigmap_{i+1}}}
\def\reverse{{\sigma_{i+1},\sigma_i,\sigma_{i-1}}}
\def\reversep{{\sigmap_{i+1},\sigmap_i,\sigmap_{i-1}}}
\def\triminus{{-\sigma_{i-1},-\sigma_i,-\sigma_{i+1}}}
\def\triminusp{{-\sigmap_{i-1},-\sigmap_i,-\sigmap_{i+1}}}
\def\tanhk{\kappa}
\def\sign{{\rm sign}}
\draft
\title{An Objective Definition of Damage Spreading - Application to
       Directed Percolation}
\author{Haye Hinrichsen$^{1}$, 
        Joshua S. Weitz$^{1,2}$ 
        and Eytan Domany$^1$}
\address{$^1$Department of Physics of Complex Systems,
         Weizmann Institute, Rehovot 76100, Israel}
\address{$^2$Princeton University, 415 Edwards Hall, USA}
\maketitle
\begin{abstract}
We present a general definition of damage spreading in a pair of
models. Using this general framework, one can define damage spreading in an 
objective manner, that does not depend on the particular dynamic procedure
that is being used. The formalism is applied to the Domany-Kinzel cellular
automaton in one dimension; the active phase of this model is shown to consist
of three sub-phases, characterized by different damage-spreading properties. 
\end{abstract}
\pacs{{\bf PACS numbers:} \ 05.70.Ln, 64.60.Ak, 64.60.Ht, 89.80.+h\\
      {\bf Key words:} \hspace{7mm} damage spreading, directed percolation}
%
% explanation of PACS numbers:
% 05.70.Ln:	Nonequilibrium and irreversible processes
% 64.60.Ak:	order-disorder phase transitions / percolation transitions
% 64.60.Ht:	dynamic critical phenomena
% 89.80.+h:	computer science
%
\renewcommand{\theparagraph}{\Alph{paragraph}}
%
%
%
%%%%%%%%%%%%%%%%%%%%%%%%%%%%%%%%%%%%%%%%%%%%%%%%%%%%%%%%%%%%%%%%%%%%%%%
\section{Introduction}
%%%%%%%%%%%%%%%%%%%%%%%%%%%%%%%%%%%%%%%%%%%%%%%%%%%%%%%%%%%%%%%%%%%%%%%
%
The concept of {\it damage spreading} was introduced in the
context of biologically motivated dynamical
systems by Stuart Kauffman\cite{Kauf}.
The question posed is whether the phase-space trajectories of two slightly
different copies of a dynamic system, subjected to the same thermal
noise, will stay close (or even merge) at long times or, alternatively,
will they diverge? Damage spreading
made its first appearance in the physics literature
in the mid eighties~\cite{Creutz,Stanley,DerWeis},
and attracted considerable interest and attention.
The main reason behind this initial enthusiasm was the hope that
damage may spread (indicating chaotic behavior)
in some regions of a system's parameter space and disappear
or {\it heal} elsewhere. This possibility intrigued researchers,
since if indeed realized, it would have indicated the
existence of different {\it dynamic phases} in
various complex systems (such as spin-glasses)
\cite{DerWeis}. The initial enthusiasm
concerning damage spreading has abated during subsequent years; the main
reason being an apparent lack of an {\it objective, observer-independent
measure} of whether damage does or does not spread in a given system.
Even for relatively simple models, such as the two dimensional
ferromagnetic Ising model, different results were obtained
when heat bath or Metropolis dynamics were used\cite{Mariz,JandArc}.
Both these dynamic procedures are phenomenological (since they satisfy
detailed balance, they can be used to generate equilibrium ensembles) and
the two are equally legitimate to mimic the temporal evolution of a system
in contact with a thermal reservoir.
If  spreading or healing of damage were to indicate some intrinsic
property of the system, one would not expect the result to depend on the
details of exactly which phenomenological procedure was used to generate
its dynamics.

The purpose of this communication is to pose the ``right'' question; i.e.
one which has a well defined objective answer. The essence of the argument
is to consider the entire family of dynamic procedures that are consistent
with the physically dictated constraints of the problem. For any particular
system one of {\it three} possibilities may hold:
\begin{enumerate}
\item
Damage is spreading for every member of the family of dynamic procedures
\item
Damage heals for every member of this family
\item
Damage spreads for a subset of the possible dynamic procedures, and heals
for the complementing subset.
\end{enumerate}
Hence the only question regarding damage spreading
that has an unambiguous, observer-independent 
answer is: to which of these three classes
a particular system belongs?

\begin{figure}
\epsfxsize=90mm
\epsffile[80 40 720 550]{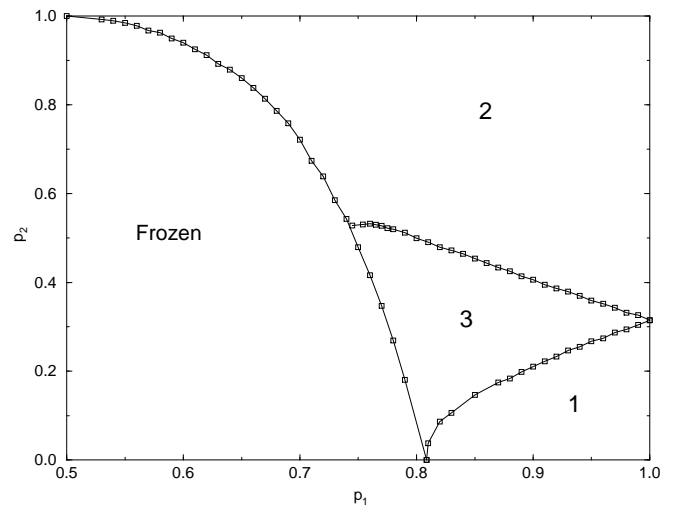}
\caption{Phase diagram of the Domany-Kinzel automaton. The active, percolating
phase consists of three sub-phases; each is numbered according to the
damage spreading class to which it belongs (see text).}
\label{Figure1}
\end{figure}

To demonstrate the general concept introduced here we studied the simplest
dynamic model in which damage spreading has been observed, the one-dimensional
Domany-Kinzel (DK) cellular automaton \cite{DK},
for which we found the phase diagram presented in Fig.~\ref{Figure1}.

The DK automaton
is a two-parameter model whose temporal evolution contains, as special cases,
the bond and site directed percolation problems.
The main point made by DK was {\it universality:} namely,
that the entire family of observed transitions of the one-dimensional
cellular automaton is in the directed percolation universality class.
%(except a special point, for which an exact solution was presented).
DK identified two phases; a ``dry'' or ``frozen''
phase, in which all initial conditions
evolve to the absorbing state, and an ``active'' or percolating phase.
Some years later Martins et al \cite{Martins1}
discovered that in a certain region of
the active phase damage spreads, and it heals elsewhere. More detailed
investigations, using simulations \cite{ZP} - \cite{GrJSP79}
 as well as analytic (mean field)
approximations \cite{Rieger}, \cite{KS} - \cite{Tom}
confirmed the existence of this ``chaotic phase''. Its
boundary, however, was shown
\cite{KS} to depend on the manner in which the dynamic
procedure of the underlying DK model is carried out, while
the evolution of a single replica is
completely insensitive to  the
dynamic procedure.
This prompted Grassberger \cite{GrJSP79}
to observe that
``it is misleading to speak of different phases in the DK
automaton...instead these are different phases
for very specific algorithms for simulating
{\it pairs} of such automata''. This observation is the precise analog of
the
problematic nature of viewing DS as a manifestation of a dynamic
transition in spin models, where, as mentioned above, it was well
known that different dynamics that yield identical equilibrium properties can
give rise to different results for damage spreading. Thus, again, DS becomes
a ``subjective'' concept, which is devoid of well defined meaning for the DK
model, whose phases should be determined by the properties of a
{\it single} evolving system.

The main purpose of
this paper is to point out that if one defines the most general family of
dynamic rules that are consistent with the physics of the problem being
studied (Sec II), DS has an objective, 
observer-independent meaning. Past work on DS in the DK model is reviewed
in Sec III and in Sec IV
the existence
of the three well defined distinct phases described in the Introduction 
is established for the DK model by numerical simulations and analytical
arguments.

We also tested and confirmed a recent conjecture of
Grassberger, to the effect that the damage spreading transition
is in the directed percolation universality class \cite{GrJSP79}.
Analytical support for this conjecture came so far
from approximate mean-field arguments \cite{Bag95} and an exact statement
first made by Kohring and Schreckenberg \cite{KS},
who noted that on the $p_2=0$ line
the dynamics of damage spreading in the DK automaton is precisely identical
to the evolution of the DK automaton itself, and hence on this line DS
is trivially in the DP universality class. This being a rather special line,
it is of interest to try to establish such precise mapping of DS to DP
also elsewhere in the $p_1 - p_2$ plane. 
In Sec IV A we present such an extension.

%
%
%
%%%%%%%%%%%%%%%%%%%%%%%%%%%%%%%%%%%%%%%%%%%%%%%%%%%%%%%%%%%%%%%%%%%%%%%%%%%
\section{Damage Spreading - General Formalism}
%%%%%%%%%%%%%%%%%%%%%%%%%%%%%%%%%%%%%%%%%%%%%%%%%%%%%%%%%%%%%%%%%%%%%%%%%%%
%
%
%
\subsection{Rules for Legitimate Damage Spreading Procedures}
%_______________________________________________________________________
%
%
\label{TruePhases}
We turn now to present our arguments for the possibility of defining
an observer-independent measurement of damage spreading. By this we
do not mean that DS is reflected in the dynamic behavior of a single
system, so that Grassberger's observation still holds; DS is a property
of a {\it pair} of
automata.\footnote{In this sense DS, as defined for stochastic
dynamics, differs from dynamics in deterministic nonlinear systems. In that
case one can find signatures of chaotic behavior in following the phase-space
trajectory of a single evolving system. Divergence of
two initially neighboring trajectories (indicating the existence of a positive
Lyapunov exponent) is a computationally feasible tool to ascertain the chaotic
nature of a single system's trajectory, but is not essential to consider two
replicas in order to {\it define} chaos.} It is possible, however,
to address the lack of objectivity implicit in one's freedom to choose
the precise algorithm that is used for the evolution of the pair of replicas.
If every observer can pick his favorite dynamic rule, get results (on DS)
that depend on the rule used, while no measurement done
on an evolving single system can differentiate between the rules - indeed
it appears contradictory
to claim that DS reflects ``phases'' of the model that is being
investigated. Nevertheless such phases can be defined in a precise way.

To overcome this apparent paradox we
formulate quite general and physically motivated restrictions on the
possible dynamic rules that one can use for studying DS. By ``physical'' we
mean that the restrictions are dictated by the dynamics of the single evolving
system. The restrictions are as follows:
\begin{enumerate}
\item
The dynamic rules for the evolution of the pair of replicas are such
that the evolution of a single replica is according to its ``natural''
dynamics.
\item
The transition probability matrix for a site $i$
for the pair of replicas can depend only
on those sites that affect the evolution of site $i$ under the
dynamic rules of a single system.
\item
The rules that govern evolution for the pair do not break any of the
symmetries of the single-replica dynamics.
\end{enumerate}
The first restriction  
simply means that the fact that we are watching two systems
evolving in parallel should not affect the behavior of any one of them.
The second constraint means that if the evolution of site $i$ is affected,
say, only by the states of its nearest neighbors, the relative states
taken on site $i$ by the two replicas should not feel longer range interactions.
For example, if site $i$ and all its neighbors $j$ are in the same state in
the two replicas, we do not expect damage to be generated at $i$ by a damaged
site which is far away (i.e. not one of the neighbors of $i$).
The third rule implies, for example,
that if there is a left-right symmetry in the evolution
of a single system the same must hold for the pair of replicas.

Clearly, the subjectivity in defining the damage spreading procedure that was
described above has now been shifted to this point - to 
selecting the restrictions that
define which DS procedure is ``legitimate''. We do believe that there is 
much less arbitrariness, however, in this kind of subjectivity 
than what was done before - 
choosing, at random, one out of a continuum of physically equivalent procedures.  
\label{Formalism}
\subsection{Transfer matrix formalism}
%-------------------------------------
%
%
We now introduce a general formalism for damage spreading problems
in one-dimensional models with parallel updates\footnote{
This is suitable for various cellular automata. For systems that satisfy
detailed balance one must not update interacting spins
simultaneously. Generalization of our formalism to random sequential update is
straightforward.}.
Consider a one-dimensional spin chain with $L$ sites and certain
boundary conditions to be specified. At each site $j=1,\ldots,L$
a local spin variable $\sigma_j=1,\ldots,s$
is placed which can be in $\statenumber$
different states. We denote by $P_t(\sigmaconf)$ the probability
to find the system at time $t$ in the state $\{\sigmaconf\}$
which is positive and normalized by
\begin{equation}
\label{Normalization}
\sum_\sigmaconf P_t \, (\sigmaconf) \;=\; 1\,.
\end{equation}
The time evolution of a system with parallel updates is discrete and
can be described by a transfer matrix $\transfermat$,
whose element $\transfermat_\sigmaconf^\tauconf$ is the probability
of making a transition from state $\{\sigmaconf\}$ at time $t$ to
state $\{\tauconf\}$ at time $t+1$. The corresponding discrete
time master equation has the form
\begin{equation}
\label{TotalUpdate}
P_{t+1}(\tauconf) \;=\; \sum_\sigmaconf
\transfermat_\sigmaconf^\tauconf\,\, P_t\, (\sigmaconf)
\end{equation}
The conservation of probability (\ref{Normalization})
implies that
\begin{equation}
\label{ProbabilityConservation}
\sum_\tauconf \transfermat_\sigmaconf^\tauconf \;=\; 1
\end{equation}
for all configurations $\{\sigmaconf\}$, which means that
the elements in each column of the transfer matrix add up to one.

In what follows we assume that each site is updated simultaneously$^2$
and that an update at some site $j$ depends locally on $n$ sites
of the previous configuration, say $\{\sigstate\}$ where
$m=\frac{n-1}{2}$. This means that the transfer matrix factorizes:
\begin{equation}
\transfermat_\sigmaconf^\tauconf \;=\;
\prod_{j=1}^{L}\, T_\sigstate^{\tau_j}\,.
\end{equation}
Each factor $T_\sigstate^{\tau_j}$ conserves probability separately:
\begin{equation}
\sum_{\tau_j=1}^s\,T_\sigstate^{\tau_j} \;=\; 1\,.
\end{equation}
We now consider two replicas $S$ and $S^\prime$ of the same system.
Denote by $\{\sigmaconf\}$ and $\{\tauconf\}$ the states of $S$
and by $\{\sigmapconf\}$ and $\{\taupconf\}$ those of $S^\prime$.
In order to define simultaneous temporal evolution of these systems,
one has to generalize the transfer matrix:
\begin{eqnarray}
\transfermat_\sigmaconf^\tauconf & \rightarrow &
\transfermat_\doublesigconf^\doubletauconf
\nonumber \\
T_\sigstate^{\tau_j} & \rightarrow & T_\sigdouble^\taudouble
\nonumber
\end{eqnarray}
The transfer matrix for the total system $(S,S^\prime)$ is restricted
by the requirement that each replica should evolve as before, i.e.
integrating out the degrees of freedom of one of the replicas results
in the previous transfer matrix of the other replica:
\begin{eqnarray}
\label{RestrictionOne}
\sum_{\tau_j=1}^\statenumber
T_\sigdouble^\taudouble &=& T_\sigpstate^{\taup_j} \\
\label{RestrictionTwo}
\sum_{\taup_j=1}^\statenumber
T_\sigdouble^\taudouble &=& T_\sigstate^{\tau_j}
\end{eqnarray}
These restrictions already imply probability conservation for the
total system:
\begin{equation}
\label{Automatic}
\sum_{\taudouble=1}^\statenumber
T_\sigdouble^\taudouble \;=\; 1
\end{equation}
In order to study damage spreading, a further restriction is imposed:
Once both replicas reach the same state (no damage), their temporal
evolution is identical:
\begin{equation}
\label{RestrictionThree}
T_\sigidentical^\taudouble \;=\; T_\sigstate^{\tau_j} \,
\delta^{\taudouble}
\end{equation}
We define damage as the Hamming distance between the states
$\{\sigmaconf\}$ and $\{\sigmapconf\}$ at any given time, that is
the fraction of sites for which $\sigma_j \neq \sigmap_j$:
\begin{equation}
\label{HammingDistance}
\Delta \;=\;  \frac{1}{L} \sum_{j=1}^L \,\Delta_j
 \;=\; \frac{1}{L} \sum_{j=1}^L \, (1-\delta_{\sigma_j,\sigmap_j})
\end{equation}
%
%
%\subsection{Independent degrees of freedom of the transfer matrix}
%-----------------------------------------------------------------
%
The restrictions (\ref{RestrictionOne}), (\ref{RestrictionTwo}),
and (\ref{RestrictionThree}) impose dependences among the $s^{2n+2}$ matrix
elements $T_\sigdouble^\taudouble$. The number of independent
degrees of freedom can be counted as follows.
First notice that because of eq. (\ref{Automatic}) only $2s-1$
of the $2s$ equations in (\ref{RestrictionOne}) and (\ref{RestrictionTwo})
are independent. Thus, even though
 for any given initial configuration $\{\sigdouble\}$
the number of different final configurations is $s^2$, the
number of independent matrix elements is $s^2-(2s-1)=(s-1)^2$.
On the other hand eq. (\ref{RestrictionThree})
implies that for $s^n$ of $s^{2n}$ possible initial configurations
$\{\sigdouble\}$ the matrix elements are already defined. Furthermore
the whole system is symmetric under exchange of the replicas which
gives another factor $\frac12$. Thus the total number of independent
degrees of freedom of the transfer matrix is
\begin{equation}
\label{DegreesFreedom}
\frac12\,(s^{2n}-s^n)\,(s-1)^2
\end{equation}
\subsection{Algorithmic implementation}
%-----------------------------------------
%
In a numerical simulation the temporal evolution described by the transfer
matrix $\transfermat$ can be realized as follows. At each time step all
sites are updated independently, i.e. $\sigma_j$ is replaced by a new
value $\tau_j$ according to probabilistic rules which depend only on
the previous configuration $\sigstate$. Since
in this paper
we discuss only two-state models, let us from now on restrict
our attention to the
case $s=2$, with $\sigma_j=0,1$
(the generalization to $s>2$ is straightforward).

We introduce a stochastic binary variable $r_\sigstate$, 
that denotes the value assigned to site $j$ in one
update, given the state of its neighborhood $\sigstate$:
\begin{equation}
\tau_j \;:=\; r_\sigstate
\label{taur}
\end{equation}
The numbers $r_\sigstate$ are generated probabilistically in some procedure,
such that updates at different sites or different times are uncorrelated.
Furthermore, their expectation value, averaged over many realizations of
random numbers, is given by the corresponding matrix element of the
transfer matrix:
\begin{equation}
\label{OnePointFunction}
\langle r_\sigstate \rangle \;=\; T_\sigstate^{\tau_j=1}
\end{equation}
Usually this procedure is implemented by generating
a random number $z$ from a uniform distribution in the interval
$0<z<1$,
and comparing it with the transition probability $T_\sigstate^{\tau_j=1}$,
to get the assigned value of  $r_\sigstate$:
\begin{equation}
r_\sigstate \;=\; \left\{
\begin{array}{ll}
1 & \mbox{if } \,z<T_\sigstate^{\tau_j=1}\\
0 & \mbox{otherwise}
\end{array}
\right.
\label{rz}
\end{equation}
This
prescription is not unique, however, and there are many other possibilities to
select $r_\sigstate$ in a manner that satisfies eq. (\ref{OnePointFunction}).
This constraint
specifies the temporal evolution of a {\it single}
system. The existence of
correlations between any two random variables $r_\sigstate$ and $r_\sigpstate$
will play no role in the evolution of
a single system, since only one of the two will be used for
any given  update.
The situation is different, however, in damage spreading problems in which
the two different replicas of the system are evolving simultaneously.
The updated state of the pair of replicas is governed by the {\it joint}
transition probabilities
\begin{equation}
{\rm Prob}(\tau_j=r_\sigma,\tau_j^\prime=r_{\sigma^\prime})=
T_\sigdouble^\taudouble
\label{myeq}
\end{equation}
Therefore
correlations between the
two random variables $r_\sigma,r_{\sigma^\prime}$
will influence the temporal evolution
of the combined system and, therewith, the properties of damage spreading.
Such correlations are contained in
the two-point correlation functions of the random variables,
which are given by the
elements of the combined transfer matrix for both replicas:
\begin{equation}
\langle r_\sigstate\,\,r_\sigpstate \rangle \;=\;
T_\sigdouble^{\tau_j=1\,; \,\tau^\prime_j=1}
\end{equation}
In this formalism the restriction
(\ref{RestrictionOne})-(\ref{RestrictionTwo}), stating that each
replica separately evolves in the same way as the original system, is
satisfied automatically. The same applies to the second restriction
(\ref{RestrictionThree})
which ensures that in case of `no damage' both systems evolve
in parallel.

Obviously, the number of degrees of freedom specifying damage spreading
is just the number of two-point correlation functions. In the case of two-state
models there are $\frac12 2^n(2^n-1)$ such correlations, which agrees
with the number of degrees of freedom counted in eq. (\ref{DegreesFreedom}).

Three-point correlations do not affect the evolution of a pair
of replicas,
because in each update only two random variables, $r_\sigstate$ and
$r_\sigpstate$, are used. However, they would start to play a role
in damage spreading problems with three replicas. Generally, $k$-point
correlations will be felt in systems consisting of at least $k$ replicas.
%
%
%
%
%%%%%%%%%%%%%%%%%%%%%%%%%%%%%%%%%%%%%%%%%%%%%%%%%%%%%%%%%%%%%%%%%%%%%%%%%%%
\section{Damage Spreading in the DK model: a brief review}
%%%%%%%%%%%%%%%%%%%%%%%%%%%%%%%%%%%%%%%%%%%%%%%%%%%%%%%%%%%%%%%%%%%%%%%%%%%
%
%
In this section we review briefly past work on damage spreading in the DK
automaton. We emphasize the manner in which DS was calculated by various
authors, and the manner in which different ways of defining DS are embedded in
the general formal framework of Sec II.

The DK automaton is defined as follows: a binary variable $\sigma_i(t)=0,1$
characterizes the state of site $i$ at (discrete)
time $t$. $\sigma=1$ means that the site
is wet or active, whereas $\sigma = 0$ means that it is dry. The
automaton evolves by a parallel update rule, which can be stated, using
the notation of Sec II, as follows:
\begin{equation}
{\bf T}_{\sigma_{i-m},\dots,\sigma_{i+m}}^{\tau_i}=
{\bf T}_{\sigma_{i-1},\sigma_{i+1}}^{\tau_i}
\label{eq:DK1}
\end{equation}
\begin{equation}
{\bf T}_{\sigma_{i-1},\sigma_{i+1}}^{1}=\left\{ \begin{array}{ll}
0 & \mbox{if $\sigma_{i-1}=\sigma_{i+1}=0$} \\
p_1 & \mbox{if $\sigma_{i-1} \neq \sigma_{i+1}$} \\
p_2 & \mbox{if $\sigma_{i-1}=\sigma_{i+1}=1$}
\end{array}
\right.
\label{eq:DK2}
\end{equation}
That is, the state of site $i$ at time $t+1$ depends only on the states
of its two neighbors at time $t$; only wet sites can give rise to a wet site,
with probabilities $p_1$ if one neighbor was wet and $p_2$ if both were wet.

Using the notations introduced in Sec II the transition probabilities
in the DK model are defined by the one-point expectation values
of {\it three} stochastic binary variables~\cite{Bag95}
\begin{equation}
\langle r_{01} \rangle = \langle r_{10} \rangle = p_1\,,
\hspace{10mm}
\langle r_{11}\rangle = p_2\,.
\end{equation}
This model has a dry phase and a wet phase, separated by a transition line
which has been determined with high accuracy by various numerical methods.
In spite of its simplicity, the model has not been solved exactly, except
for the special line $p_2 = 1$ \cite{DK},\cite{Essam},\cite{Dickman95}.
At all points on the phase boundary, except the special line,
the transition to the active or wet phase
is characterized by directed percolation (DP) exponents.

Damage spreading properties between two replicas are controlled by
correlations between the random variables:
\begin{eqnarray}
\langle r_{01} \, r_{11} \rangle \;=\; \langle r_{10} \,r_{11} \rangle
&=& {\tilde \alpha} \nonumber \\
\langle r_{01} \, r_{10} \rangle &=& {\tilde \beta} \\
\langle r_{01} \, r_{10} \, r_{11} \rangle &=& {\tilde \gamma} \nonumber
\end{eqnarray}
According to the arguments discussed in Sec II, only
one- and two-point functions enter the transfer matrix, which means
that ${\tilde \gamma}$
is an irrelevant parameter in the present problem.

Martins et al \cite{Martins1} were the first to address the issue of damage
spreading in the DK model. Two nearly identical initial configurations were
allowed to evolve on two replicas, using the same
random numbers for both
(the precise meaning of this statement will be explained below).
They discovered that the active phase contains
in fact two regions; one  in which damage spreads and its complement, where
it does not. The boundary between these regions
was subsequently determined with
increasing accuracy by Zebende and Penna \cite{ZP},  by Martins et al
\cite{Martins2}, Rieger et al \cite{Rieger} and Grassberger
\cite{GrJSP79}.
Independently, mean-field type approximations of varying
complexity were also used to study the DS problem \cite{Rieger,KS,Bag95,Tom}.
An interesting observation, first made by Kohring and Schreckenberg
\cite{KS}, was
to the effect that the position of the ``phase boundary'' between the DS and
non-spreading regimes depends on the manner in which damage is generated.
The original scheme of Martins et al used a single
uniformly distributed
random number $0 < z < 1$
for the
two replicas: using the definitions of Sec II this means that the
 choice
\begin{equation}
r_{01}=r_{10}=\theta ( p_1 - z)\,, \qquad \qquad
r_{11}=\theta(p_2 - z)
\label{eq:1rand}
\end{equation}
was made, which can be expressed as
\begin{eqnarray}
{\tilde \beta} & = \langle r_{01}r_{10} \rangle &= p_1 \nonumber \\
{\tilde \alpha} &= \langle r_{01}r_{11}\rangle &={\rm Min}(p_1,p_2)
\end{eqnarray}
The dynamical process is generated by setting
\begin{eqnarray}
\sigma_i(t+1)&=&\tau_i=r_{\sigma_{i-1}\sigma_{i+1}}\nonumber\\
\sigma_i^\prime(t+1)&=&\tau_i^\prime=
r_{\sigma_{i-1}^\prime \sigma_{i+1}^\prime}
\nonumber
\end{eqnarray}
For reasons that will be evident shortly, we call
this dynamic process, that uses
a single random number, {\it maximally correlated}.

Kohring and Schreckenberg
recognized the fact that one could, in principle, use two different
random numbers to determine $\tau_i$ and $\tau_i^\prime $, if at least
one of the two neighbor sites was damaged at time $t$. In fact they studied DS
using two different random numbers $z_{01}$ and $z_{11}$, their DS procedure
has two {\it fully correlated} binary variables ($r_{01}$ and $r_{10}$) and
two {\it uncorrelated} ones ($r_{01}$ and $r_{11}$):
\begin{equation}
r_{01}=r_{10}=\theta(p_1 -z_{01})\,,   \qquad \qquad
r_{11}= \theta(p_2 - z_{11})
\label{eq:2rand}
\end{equation}
the correlations being
\begin{eqnarray}
{\tilde \beta} &= \langle r_{01}~r_{10} \rangle &=p_1 \nonumber \\
{\tilde \alpha} &= \langle r_{01}~r_{11}\rangle &= \langle r_{01} \rangle
\langle r_{11} \rangle=p_1~p_2
\label{eq:KScorr}
\end{eqnarray}
The dynamics
generated by using
on the first replica $\tau_i=r_{\sigma_{i-1}\sigma_{i+1}}$
and $\tau_i^\prime=r_{\sigma_{i-1}^\prime\sigma_{i+1}^\prime}$ on the second
gave rise to a shift of the original ``phase boundary''
(as obtained with a single random number,
eq. (\ref{eq:1rand})).
As discussed in Sec II, the evolution of a single replica is
completely insensitive to whether one or two random variables are used in the
dynamic procedure,
which prompted Grassberger \cite{GrJSP79}
to make his observation quoted in the Introduction.

Finally we note that Grassberger has formulated recently
\cite{GrJSP79} a conjecture,
which is a natural extension of previous statements
\cite{Janssen,Gr82,DK} regarding universality
of directed percolation transitions for models with non-symmetric absorbing
states \cite{ParkPRL}.
According to this conjecture
damage-spreading transitions should be in the universality
class of directed percolation\cite{GrJSP79}, provided some general conditions
are satisfied. The DK model is a natural candidate to test this conjecture
because of its simplicity, ease to simulate and our precise knowledge of the
existence of a DS transition and its location.
Grassberger presented numerical evidence for his conjecture, which we also
confirmed and extended.
We also show below that in a region of the $p_1,p_2$ plane one can map
DS exactly to the DK model and hence onto DP. This result is an extension
of a
statement
first made by Kohring and Schreckenberg \cite{KS},
who noted that such a mapping holds on the $p_2=0$ line.

%
%%%%%%%%%%%%%%%%%%%%%%%%%%%%%%%%%%%%%%%%%%%%%%%%%%%%%%%%%%%%%%%%%%%%%%%%%%%
\section{True phases in the DK model}
%%%%%%%%%%%%%%%%%%%%%%%%%%%%%%%%%%%%%%%%%%%%%%%%%%%%%%%%%%%%%%%%%%%%%%%%%%%
%
%
As discussed in Sec III, the most general dynamic
rule that can be defined for two replicas of the DK automaton, in accordance
with these constraints, has two degrees of freedom or parameters, ${\tilde
 \alpha}$
and ${\tilde \beta}$. As it turns out (see Appendix~\ref{AppendixRestrictions}),
the possible values that ${\tilde \alpha}$ and ${\tilde \beta}$
can take are restricted by requiring that
all transition rates in the transfer matrix have to be positive.
For any value of $p_1,p_2$, the range of allowed values of the
parameters ${\tilde \alpha}$ and ${\tilde \beta}$ is given by
\begin{eqnarray}
\label{ParameterRange}
&&\max(0,p_1+p_2-1) \;\leq\; {\tilde \alpha} \;\leq\; \min(p_1,p_2)\nonumber\\
&&\max(0,\,2{\tilde \alpha}-p_2,\,2p_1-1,\,2p_1-1-2{\tilde \alpha}+p_2)
\;\leq\; {\tilde \beta} \;\leq\; p_1 \nonumber \\
\end{eqnarray}
There are three important special cases, namely those of
\begin{itemize}
\item maximal correlations: \   ${\tilde \alpha}=\min(p1,p2),\,\, {\tilde
 \beta}=p_1$
\item no correlations: \ \ \ \ \ \ \ \ \ \ ${\tilde \alpha}=p_1p_2,\,\,{\tilde
 \beta}=p_1^2$
\item minimal correlations: \ \ \parbox[t]{40mm}{${\tilde \alpha}=p_1+p_2-1$,\\
				      ${\tilde \beta}=2p_1-1$}

\end{itemize}
In the case of minimal correlations, the values listed above hold
only in the
region $2p_1+p_2>2$ (see Appendix~\ref{AppendixRestrictions}).
\subsection{Exact results}
%---------------------------------------------
%
We turn now to show that for $p_2/2 \leq p_1 \leq 1-p_2/2$
the damage spreading
process can be mapped exactly onto a directed percolation process.
Kohring and Schreckenberg \cite{KS} have shown that such a mapping holds
on the {\it line} $p_2 = 0$. Clearly, their choice of parameters
(\ref{eq:KScorr}) is a particular case of
our damage spreading procedure, which is the most general one
that satisfies rules 1 - 3 listed above. Therefore
we find a wider ({\it two-dimensional}) region
in the $p_1,p_2$ plane in which such a mapping is possible. To see this,
let $\Delta_i=1-\delta_{\sigma_i,\sigma^\prime_i}$ be the damage at
site $i$. By
$P_D(\Delta_i=1\,|\, \sigma_{i-1}\sigma_{i+1};
\sigma^\prime_{i-1}\sigma^\prime_{i+1})$
we denote the probability to generate a damaged site for a given initial
configuration in a particular update. These probabilities
are listed in Table I, in which we introduced
for brevity the notation
\begin{equation}
X=2 p_1 - 2{\tilde \beta} \qquad \qquad Y=p_1+p_2-2{\tilde \alpha}
\label{eq:XY}
\end{equation}
In general, the probability for generating damage on site $i$ depends
on the previous states of both
replicas, i.e. on $(\sigma_{i-1}\sigma_{i+1};
\sigma^\prime_{i-1}\sigma^\prime_{i+1})$; knowledge of $\Delta_{i-1}$ and
 $\Delta_{i+1}$ does not suffice to determine  $\Delta_i$ at the next time step.
Thus damage spreading itself cannot be seen as an independent
process. We may, however, pose the following question:
under which conditions will damage
spread as if it were generated by an independent process? That is,
when do we have
\begin{equation}
\label{IndependentProcess}
P_D(\Delta_i\, | \, \sigma_{i-1}\sigma_{i+1};
\sigma^\prime_{i-1}\sigma^\prime_{i+1})
\;=\; P_D(\Delta_i \, | \, \Delta_{i-1},\Delta_{i+1})
\end{equation}
In order to satisfy this condition, any two entries
in Table I,
that correspond to the same initial damage $\{\Delta_{i-1}\Delta_{i+1}\}$,
should be equal.
For example all four initial configurations
\begin{eqnarray}
\{\sigma_{i-1}\sigma_{i+1}&;&\sigma^\prime_{i-1}\sigma^\prime_{i+1}\} 
\nonumber \\
&=& (\{11;00\},\{10;01\},\{01;10\},\{00;11\}) \nonumber
\end{eqnarray}
have the same initial
damage $\{\Delta_{i-1},\Delta_{i+1}\}=\{1,1\}$.
In order to satisfy eq. (\ref{IndependentProcess}), the four entries
$(p_2,X,X,p_2)$ must have the same value, i.e.
we must have $p_2=X$. A similar consideration leads to
the condition $Y=p_1$; that is, we must have
\begin{eqnarray}
&&P_D(1|00)=0 \nonumber \\
&&P_D(1|01)=P_D(1|10)=p_1=Y \\
&&P_D(1|11)=p_2 =X\nonumber
\end{eqnarray}
Note that these are precisely the update rules of the DK process.
Using the definitions~(\ref{eq:XY}), we see that
the correlations must satisfy
\begin{equation}
{\tilde \alpha}=\frac{p_2}{2}\,, \qquad
{\tilde \beta}=p_1-\frac{p_2}{2}\,.
\label{eq:albe}
\end{equation}
Since the correlation parameters are restricted by eq.~(\ref{ParameterRange}),
the allowed range for $p_1$ and $p_2$
in which these conditions  can hold
is a triangle in the phase diagram:
\begin{equation}
p_2/2 \,\leq\, p_1 \,\leq\, 1-p_2/2
\label{eq:triangle}
\end{equation}
To summarize: we have proved that within this triangle we can find
correlations ${\tilde \alpha},{\tilde \beta}$ such that the damage spreading
 process
follows the dynamical rules of a single DK automaton.

\begin{table}
\label{TableGenerateDamage}
\begin{center}~
\begin{tabular}{|c||c|c|c|c|}
%\hline
&\multicolumn{4}{c|}{$\sigma^\prime_{i-1}, \sigma^\prime_{i+1}$} \\
\hline
$\sigma_{i-1},\sigma_{i+1}$ & \bf 00 & \bf 01 & \bf 10 & \bf 11 \\
\hline \hline
\bf 00 & $0$ & $p_1$ & $p_1$ & $p_2$ \\
%\hline
\bf 01 & $p_1$ & $0$ & $X$ & $Y$ \\
%\hline
\bf 10 & $p_1$ & $X$ & $0$ & $Y$ \\
%\hline
\bf 11 & $p_2$ & $Y$ & $Y$ & $0$ \\
%\hline
\end{tabular}
\end{center}
\caption{Probabilities
$P_D(\Delta_i=1\,|\, \sigma_{i-1}\sigma_{i+1};
\sigma^\prime_{i-1}\sigma^\prime_{i+1})$
for the generation of damage in the DK model.
$X$ and $Y$ are defined in eq. (26)}
\end{table}

Say we have a
line in the $(p_1,p_2)$ plane that lies within this region.
For every point $(p_1^*,p_2^*)$
on this line we can find ${\tilde \alpha},{\tilde \beta}$ values for which
DS evolves precisely like a DK automaton with parameters $(p_1^*,p_2^*)$.
Since part of the transition line of the DK model (from dry to wet phase)
lies in the triangle~(\ref{eq:triangle}), on any trajectory that crosses
this part of the phase boundary we will observe a damage spreading transition
precisely at the DP transition and with DP exponents (provided we chose
${\tilde \alpha},{\tilde \beta}$ according to eq.~(\ref{eq:albe}).) In
 particular, this holds
for the line $p_2=0$, as discovered in \cite{KS}; note that
for $p_2=0$ their choice of
correlations, eq.~(\ref{eq:KScorr}) precisely satisfy eq.~(\ref{eq:albe}).

\subsection{Results from comparing probabilities}
%-----------------------------------------------
\label{AppendixCompare}
Other useful results can be obtained by comparing probability
tables of different pairs of automata. The basic idea is
that by increasing (decreasing) {\em all} probabilities
in Table I,
damage spreading will be more (less) likely. More
precisely, if a pair of DK automata described
by parameters $p_1^*,p_2^*,\tilde{\alpha}^*,\tilde{\beta}^*$
exhibits damage spreading, we expect that any other pair of automata
with parameters $p_1,p_2,\tilde{\alpha},\tilde{\beta}$ satisfying
\begin{eqnarray}
\label{MoreDamage}
p_1^* &\leq& p_1 \nonumber \\
p_2^* &\leq& p_2 \nonumber \\
p_1^*+p_2^*-2\tilde{\alpha}^* &\leq& p_1+p_2-2\tilde{\alpha}  \\
2p_1^*-2\tilde{\beta}^* &\leq& 2p_1-2\tilde{\beta} \nonumber
\end{eqnarray}
exhibits damage spreading as well. Vice versa, if  damage heals
in a pair of automata described by
$p_1^*,p_2^*,\tilde{\alpha}^*,\tilde{\beta}^*$,
then for any other pair with $p_1,p_2,\tilde{\alpha},\tilde{\beta}$ obeying
\begin{eqnarray}
\label{LessDamage}
p_1^* &\geq& p_1 \nonumber \\
p_2^* &\geq& p_2 \nonumber \\
p_1^*+p_2^*-2\tilde{\alpha}^* &\geq& p_1+p_2-2\tilde{\alpha}  \\
2p_1^*-2\tilde{\beta}^* &\geq& 2p_1-2\tilde{\beta} \nonumber
\end{eqnarray}
we expect damage to heal.

Although these statements are very plausible,
we were not able to prove them rigorously. However, we
performed various numerical tests which turned out to be
consistent with the inequalities stated above.

Because of these inequalities the boundaries between the three
regions in the phase diagram correspond to extremal correlations
${\tilde \alpha}$ and ${\tilde \beta}$. For example, if 
at a point $(p_1,p_2)$ damage spreads
in a model with maximal correlations $\tilde{\alpha}^{max}=\min(p_1,p_2)$ and
$\tilde{\beta}^{max}=p_1$, then eq. (\ref{MoreDamage}) implies that damage
spreads also for every $\tilde{\alpha}$ and $\tilde{\beta}$ in the allowed
range (\ref{ParameterRange}). This, however, 
means that the point $(p_1,p_2)$ belongs to
region 1 in the phase diagram. Therefore the phase boundary of region 1
coincides with the DS transition line for maximal correlations. Similarly
one can use eq. (\ref{LessDamage}) to show that the phase boundary between
regions 2 and 3 coincides with the DS transition line for minimal correlations.
It turns out (see Fig. \ref{Figure2})
that this line lies entirely in the region $2p_1+p_2>2$ so
that minimal correlations are well defined (see Appendix~\ref{AppendixMaxMin}).

Alternatively one can compare the probabilities for generating
damage in a pair of DK automata to the probabilities of 
generating a wet site in a single DK automaton.
To this end one simply has to use the same inequalities setting
$\tilde{\alpha}^*=p_2^*/2$ and $\tilde{\beta}^*=p_1^*-p_2^*/2$.
For example, if $p_1^*$ and $p_2^*$ represent a point in the wet phase of the
DK phase diagram, then for all pairs of automata
parametrized by $p_1,p_2,{\tilde \alpha}, {\tilde \beta}$ and satisfying
\begin{eqnarray}
\label{DKMoreDamage}
p_1^* &\leq& p_1 \nonumber \\
p_2^* &\leq& p_2 \nonumber \\
p_1^* &\leq& p_1+p_2-2\tilde{\alpha}  \\
p_2^* &\leq& 2p_1-2\tilde{\beta} \nonumber
\end{eqnarray}
damage will spread.
On the other hand, if $p_1^*$ and $p_2^*$ belong
to the dry phase of the DK model then in all pairs of
automata with
\begin{eqnarray}
\label{DKLessDamage}
p_1^* &\geq& p_1 \nonumber \\
p_2^* &\geq& p_2 \nonumber \\
p_1^* &\geq& p_1+p_2-2\tilde{\alpha}  \\
p_2^* &\geq& 2p_1-2\tilde{\beta} \nonumber
\end{eqnarray}
damage does not spread.

As an illustration of eq. (\ref{DKMoreDamage})
consider the point $M_1$ in Fig. \ref{Figure2}:
Setting $p_1^*=p_1^c\approx 0.809$ and
$p_2^*=0$ we obtain the conditions
\begin{equation}
p_1^c \leq p_1 \,,
\hspace{10mm}
p_1^c \leq p_1+p_2-2\tilde{\alpha}   \,,
\hspace{10mm}
0 \leq 2p_1-2\tilde{\beta}
\end{equation}
Using the bounds~(\ref{ParameterRange}) we find from these inequalities 
that in the triangle $p_1-p_2 \geq p_1^*$
damage spreads with certainty for any $\tilde{\alpha}$ in the allowed range.
In Fig. \ref{Figure2} this triangle is indicated as a shaded region.
\subsection{Terminal points of the phase boundaries}
\begin{figure}
\epsfxsize=90mm
\epsffile[60 480 390 780]{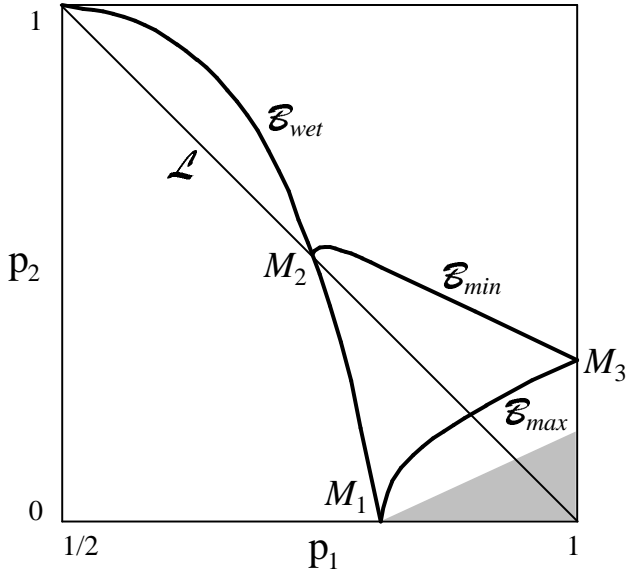}
\caption{Schematic phase diagram, displaying the special lines and points
that are referred to in the text.}
\label{Figure2}
\end{figure}

%------------------------------------------------------------
We turn now to derive, using the arguments introduced above,
a few exact results concerning
the phase boundaries for minimal and maximal
correlations. As explained above, these boundaries are the
transition lines between the damage-spreading phases 1,2 and 3 described
in the Introduction and shown on   \ref{Figure1}. In order to
make our arguments easier to follow, we present in
Fig.~\ref{Figure2} all the lines and special points that
are mentioned.

Let us consider first the case of {\it maximal correlations}.
That is, for every point ($p_1,p_2$) in the phase diagram
we assign ${\tilde \alpha}= {\rm min}(p_1,p_2)$ and ${\tilde \beta}=p_1$
and look for the boundary ${\cal B}_{max}$
between the region in which damage spreads
and the one in which it doesn't. Denote the wet-to-dry
transition line of the DK model by ${\cal B}_{wet}$
(see Fig.~\ref{Figure2}).

We now
prove that ${\cal B}_{max}$ and ${\cal B}_{wet}$ can intersect only
at the point $M_1=(p_1^c,0)$, where ${\cal B}_{wet}$
intersects  the  $p_1$ axis. On this axis maximal correlations
correspond to the choice ${\tilde \alpha}=p_2=0$ and ${\tilde \beta}=p_1$,
which also
satisfy the conditions
(\ref{eq:albe}), i.e.  the damage spreading process can be mapped
onto a single DK model. Therefore we know that a DS transition will
occur precisely at $p_1=p_1^c$, so that at all
points $(p_1>p_1^c,p_2=0)$
we must have DS. On the other hand, as we will now show, for maximal
correlations there cannot
be DS on any point on ${\cal B}_{wet}$;
%hence the boundary  ${\cal B}_{max}$ must go through $M_1$.
%In order to substantiate the last claim, 
to see this, note that (for the region of interest,
$p_1 > 1/2$)
maximal correlations imply the inequalities
\begin{eqnarray}
\label{usenext} 
p_1+p_2-2{\tilde \alpha} = |p_1-p_2| &<& p_1\\
2p_1-2{\tilde \beta}     = 0         &<& p_2\nonumber
\end{eqnarray}

For a point $(p_1,p_2)$ on  ${\cal B}_{wet}$ these inequalities, when used
together with eq. (\ref{DKLessDamage}), imply that damage does not spread.
Furthermore, since~(\ref{usenext}) are (for $p_2>0$) strict inequalities, 
 ${\cal B}_{wet}$ lies {\it inside} the no-spread region. The point 
$M_1$ is on the boundary 
between this region and a region (containing
the $p_1>p_1^c$ axis) in which damage does spread; 
Therefore $M_1$ must lie on ${\cal B}_{max}$.

We turn now to
the phase boundary ${\cal B}_{min}$ for
{\it minimal correlations}, and show that it
terminates at the point $M_2$,
where ${\cal B}_{wet}$ intersects the line  $\cal L$,
given by $p_1+p_2/2=1$.
First
notice that on $\cal L$ minimal correlations correspond to
${\tilde \alpha}=1-p_1$ and ${\tilde \beta}=2p_1-1$.
Therefore (see eq.~(\ref{eq:albe})) the
DS process is equivalent, on
$\cal L$, to  a DK model, so that as we move on $\cal L$, keeping
minimal correlations,
a DS transition occurs at the point $M_2=(p_1^*,p_2^*)$, where $\cal L$
intersects ${\cal B}_{wet}$. As before, we show next
that at all points on ${\cal B}_{wet}$ with $p_2>p_2^*$ damage doesn't spread;
hence ${\cal B}_{min}$ must pass through $M_2$. To prove the last
claim note that the segment of ${\cal B}_{wet}$ that lies above
the intersection point is in the region where
$p_1+p_2/2>1$ and this immediately leads
(for minimal correlations) to the inequalities
\begin{eqnarray}
p_1+p_2-2{\tilde \alpha} = 2-p_1-p_2 &<& p_1 \\
2p_1-2{\tilde \beta}     = 2-2p_1    &<& p_2 \nonumber 
\end{eqnarray}
According to eqs. (\ref{DKLessDamage}), this implies that
on the DK transition line in the region $p_1+p_2/2>1$ damage
does not spread. Therefore $M_2$ is the terminal point of ${\cal B}_{min}$.

Having located the endpoints $M_1$ and $M_2$,
we now turn to the opposite end of  the lines
${\cal B}_{max}$ and  ${\cal B}_{min}$.
Note that for $p_1=1$ the bounds~(\ref{ParameterRange}) collapse to
${\tilde \alpha}=p_2, {\tilde \beta}=1$, i.e. maximal and minimal
correlations are identical
and hence ${\cal B}_{max}$ and  ${\cal B}_{min}$ meet
at some point $M_3$ on the $p_1=1$ line.
The three special ``multicritical'' points discussed above determine
the topology of the phase diagram for DS. In order
to obtain  high precision quantitative information about the location
of the transition lines we performed numerical studies of
damage spreading in the DK model.
\subsection{Numerical results}
%------------------------------------------------------------
%
In order to obtain  accurate numerical
estimates for the critical parameters of models with absorbing
states one usually has to let the system evolve for extremely long
times~\cite{Martins1}.
Grassberger overcame the difficulty posed by long transients and obtained
good statistics
by simulating $n$ replicas of the
same system in parallel, using simple bit manipulations on computers
with unsigned words of length $n$~\cite{GrJSP79}.
Using this multi-spin encoding method
he measured the decay in damage on a one-million site chain, allowing
it to evolve for
hundreds of thousands of time steps. Because of the improved statistics
he was able to determine the critical exponents for damage spreading at a
particular transition point with high accuracy.

	Another method to determine the critical point efficiently is the
so-called gradient method which was introduced by Zebende and
Penna~\cite{ZP}. In this method a gradient in $p_1$ and $p_2$ is
arranged along the chain. The values
of the parameters at the two  end-points of the chain are
chosen to be in different phases, i.e. on different sides of the
transition point. This allows the critical point to be determined
by measuring the average location of the boundary of the active (damaged)
cluster.

	In the present work we used a
combination of multi-spin encoding and the gradient
method. In combining these methods, a number of problems emerged which we
solved as follows:

\begin{enumerate}
\item
	In order to measure the damage spreading transition point, one has
	to find the first position (approaching from the non-spreading phase)
	where damage occurs. Simulating $n=64$ lattices in parallel, this
	has to be done for each of the $64 \cdot 63/2=2016$ pairs
        of replicas. To do this one has
	to set up a $64 \times 64$ table in order to keep track of
	damaged pairs. Moreover,
	one has to scan the words bit by bit which makes
	it impossible to use parallel bit manipulations. The large amount of
	CPU time needed for this process usually kills the advantage one gains
	from the multi-spin encoding. In order to solve this problem, we
	used a simplified search algorithm which is based on fast bit
	operations. The price we pay is that only $\approx~75\%$ of all
	possible pairs are taken into account\footnote{In this approximation,
	a given replica is declared to be damaged at site $i$ if the majority
	of the other replicas is in a different state. Instead of keeping
	track of damaged {\it pairs of replicas}, we recognize only
	{\it single replicas} from where damage originated. This amounts in
	dropping statistically 25\% of all possible pairs.}. We proved that
	the error of this method does not bias the measurement of the
	transition point.
\item
	Zebende and Penna started each run with a single damaged seed located
	somewhere on the chain. It is not clear whether the choice
	of the location influences the results. In order to circumvent this
	problem, we used initial conditions with  randomly distributed
	damage all over the chain.
\item
	The gradient method is a finite-size simulation and therefore
	boundary conditions may play an important role. In the work of
	Zebende and Penna the boundary conditions can be understood as
	dry walls and it is not clear to what extent they affect the
	measurements. In order to minimize this effect, we created,
	on a chain of $2N$ sites,
	a gradient with reflection symmetry ($p_1(i)=p_1(2N-i)$
	and $p_1(i)=p_1(2N-i)$) and measured the boundary of the damaged
	cluster on both sides. We expect finite-size
	effects to be less important for these
        periodic boundary conditions.
\end{enumerate}

The phase diagram of the DK automaton, obtained
using the {\it multiple lattice gradient method},
is presented in Fig.~\ref{Figure1}. First, we
verified numerically the prediction that
larger correlations correspond to smaller damage and vice versa. This was done
 by scanning the $({\tilde \alpha},{\tilde \beta})$ space for various points in
 the
$(p_1,p_2)$ plane.
Next we determined the DS transition lines for minimal and maximal
correlations. Typical gradient values of $1.2 \cdot 10^{-5}$ were used
for lattice sizes $L=8192$ and upwards. A transient period of at least
$2L$ was followed by an averaging period of $L$ time steps. For $(p_1,p_2)$
near the the transition lines longer transient times were used. The terminal
points of the phase boundaries were determined with high accuracy. Using a
chain with $L=16384$ sites, gradients down to $1.22 \cdot 10^{-6}$ and
transients of 231072, we measured the
following critical values at these special
points:
$p^c_1=0.8087(5)$ (on the $p_2=0$ line);
$p_2^c=0.3130(5)$ (on the $p_1=0$ line). The new triple
point was located  at $p_1^*=0.744(10), p_2^*=0.526(10)$.
This was done {\it without} using our analytic result that identified
this point as the crossing of $\cal L$ with ${\cal B}_{wet}$; the value of
$p_1^*+ p_2^*/2=1.007(15)$ agrees with the predicted value (of 1) for points
on   $\cal L$.

Measuring the density of damage along the gradient of the chain, we
could estimate the density exponent $\beta$. At the terminal points
we found $\beta=0.302(30)$ for $p_2=0$ and $\beta=0.296(30)$
on the $p_1=1$ line.
We also measured the exponent at a point $(p_1,p_2)=(0.85,0.35)$ which
lies inside phase 3. This was done by crossing the DS phase boundary
while varying the correlations ${\tilde \alpha}$
and ${\tilde \beta}$, yielding the value $\beta=0.279(10)$.  All results
are in fair agreement with the expected density exponent of
directed percolation $\beta=0.277(1)$~\cite{DickmanPRL91},\cite{EssamAdler}.

%THERE IS A CONFLICT IN NOTATIONS:
%THE DENSITY EXPONENT BETA AND ALPHA/BETA-SPACE
%
%

%
%
%
%
%%%%%%%%%%%%%%%%%%%%%%%%%%%%%%%%%%%%%%%%%%%%%%%%%%%%%%%%%%%%%%%%%%%%%%%%%%%
\section{Summary}
%%%%%%%%%%%%%%%%%%%%%%%%%%%%%%%%%%%%%%%%%%%%%%%%%%%%%%%%%%%%%%%%%%%%%%%%%%%
%
%
\label{Summary}
We have rules that a most general damage spreading procedure should satisfy.
These rules are most natural: they
ensure that the evolution of a single replica is not
affected by the fact that two replicas are evolving simultaneously; that 
the range of damage spreading does not exceed the range of interactions in
the original single model and that the two evolving replicas respect the
symmetries of the model.
These rules can be cast in a formal setting, using transfer matrices.
Using these formal definitions we were able to parametrize the most general
damage spreading procedure for any given model in terms of correlation
coefficients between various stochastic binary variables. Thus we are
considering {\it all possible damage spreading procedures} and identify
different damage spreading {\it phases} in terms of the manner in which
this complete set of procedures behaves. Three possible phases can occur;
one in which damage spreads for {\it all} allowed procedures, one in which
it does not spread for any procedure and the third, in which for some
procedures damage spreads while for others it does not. 

These ideas were implemented for the Domany-Kinzel automaton, for which
the three phases were identified, using a combination of numerical and
analytic methods. We have shown that in
an extended region of the model's parameter space damage spreading can be
mapped onto the evolution of the DK automaton itself. This observation
supports Grassberger's recent conjecture to the effect that damage spreading
is in the directed percolation universality class. This was also confirmed
by numerical tests (performed in regions where the above mentioned mapping
does not hold). 
%
%
%
%%%%%%%%%%%%%%%%%%%%%%%%%%%%%%%%%%%%%%%%%%%%%%%%%%%%%%%%%%%%%%%%%%%%%%%%%%%%
%                                 APPENDICES
%%%%%%%%%%%%%%%%%%%%%%%%%%%%%%%%%%%%%%%%%%%%%%%%%%%%%%%%%%%%%%%%%%%%%%%%%%%%
%
\appendix
\section{Generation of correlated random variables}
%--------------------------------------------------
\label{AppendixRestrictions}
In this Appendix we explain in detail how correlated random variables
$r_{01}$, $r_{10}$ and $r_{11}$,
that govern the evolution of the DK-model, can be generated.
We also prove the allowed ranges
for ${\tilde \alpha}$ and ${\tilde \beta}$, given
in eq. (\ref{ParameterRange}). Finally, we explain the manner in which
minimal correlations are given by the expression  presented in Sec. V.

Since in each update $r_{01}$, $r_{10}$, and $r_{11}$
can be either zero or one, there are eight possible combinations.
By $\pi_{r_{01},r_{10},r_{11}}$ we denote the (positive) probability
to generate the combination $\{r_{01},r_{10},r_{11}\}$.
These probabilities are normalized by:
\begin{equation}
\label{PiNorm}
\pi_{000} + \pi_{001} + \pi_{010} + \pi_{011} +
\pi_{100} + \pi_{101} + \pi_{110} + \pi_{111} \;=\; 1\,.
\end{equation}
Once the probabilities $\pi_{r_{01},r_{10},r_{11}}$ are known,
the numbers $r_{01}$, $r_{10}$, and $r_{11}$ can be generated by taking one
uniformly distributed random number $0<z<1$ and selecting
one of the eight possible outcomes according to the probabilities
$\pi_{r_{01},r_{10},r_{11}}$. 
The three-point correlation functions
can be represented in terms of the $\pi$'s: 
\begin{eqnarray}
\pi_{111}&=&\langle r_{01}r_{10}r_{11} \rangle \nonumber \\
\pi_{110}&=&\langle r_{01}r_{10}\,(1-r_{11}) \rangle \nonumber
\end{eqnarray}
Appropriate combinations of these probabilities are related to
two- and one-point functions and therewith to the entries of the
transfer matrix; for example,
\begin{equation}
\pi_{110}+\pi_{111}  \;=\; \langle r_{01}r_{10} \rangle \;=\;
T_{01;10}^{1;1} = {\tilde \alpha}
\end{equation}
Collecting all identities of this type, we obtain seven equations:
\begin{eqnarray}
\label{PiRelations}
&& \pi_{100} + \pi_{101} + \pi_{110} + \pi_{111} \;=\; p_1 \nonumber \\
&& \pi_{010} + \pi_{011} + \pi_{110} + \pi_{111} \;=\; p_1 \nonumber \\
&& \pi_{001} + \pi_{011} + \pi_{101} + \pi_{111} \;=\; p_2 \nonumber \\
&& \pi_{011} + \pi_{111} \;=\; {\tilde \alpha} \\
&& \pi_{101} + \pi_{111} \;=\; {\tilde \alpha} \nonumber \\
&& \pi_{110} + \pi_{111} \;=\; {\tilde \beta} \nonumber \\
&& \pi_{111} \;=\; {\tilde \gamma} \nonumber
\end{eqnarray}
Together with the normalization (\ref{PiNorm}) these equations
determine all probabilities $\pi_{r_{01},r_{10},r_{11}}$:
\begin{eqnarray}
\pi_{000} &=& 1-{\tilde \gamma}+{\tilde \beta}+2{\tilde \alpha}-2p_1-p_2
 \nonumber \\
\pi_{001} &=& p_2-2{\tilde \alpha}+{\tilde \gamma} \nonumber \\
\pi_{010} \;=\; \pi_{100} &=& p_1-{\tilde \alpha}-{\tilde \beta}+{\tilde
 \gamma}\\
\pi_{011} \;=\; \pi_{101} &=& {\tilde \alpha}-{\tilde \gamma} \nonumber \\
\pi_{110} &=& {\tilde \beta}-{\tilde \gamma} \nonumber \\
\pi_{111} &=& {\tilde \gamma} \nonumber
\end{eqnarray}
Since all $\pi$ have to be positive, we obtain six inequalities:
\begin{eqnarray}
\label{Restrict1}
\max(0,\,2{\tilde \alpha}-p_2) &\leq& {\tilde \gamma} \;\leq\; {\tilde \alpha}
 \\
\label{Restrict2}
\max({\tilde \gamma},\,{\tilde \gamma}-1-2{\tilde \alpha}+2p_1+p_2) &\leq&
 {\tilde \beta} \;\leq\;
{\tilde \gamma}+p_1-{\tilde \alpha}
\end{eqnarray}
For a given choice of the parameters $p_1,p_2$ these inequalities
imply restrictions on the correlation parameters ${\tilde \alpha}$, ${\tilde
 \beta}$ and ${\tilde \gamma}$.
The allowed range of these parameters can be derived as follows.
First let us consider the restrictions on ${\tilde \alpha}$.
Eq. (\ref{Restrict1}) implies that $0 \leq {\tilde \alpha} \leq p_2$ whereas
eq. (\ref{Restrict2}) leads to the condition $p_1+p_2-1\leq{\tilde \alpha}\leq
 p_1$.
Both of them can be combined by requiring
\begin{equation}
\label{AlphaRange}
\max(0,p_1+p_2-1) \leq {\tilde \alpha} \leq \min(p_1,p_2)
\end{equation}
For a given ${\tilde \alpha}$ in this interval
the maximal ranges of ${\tilde \beta}$ and ${\tilde \gamma}$ are given in eqs.
(\ref{Restrict1})-(\ref{Restrict2}). However, since we do not explicitly
use the three-point correlation parameter ${\tilde \gamma}$, we are only
 interested
in the maximal range of ${\tilde \beta}$. This range can be obtained by
 inserting
the extremal values for ${\tilde \gamma}$ into eq.~(\ref{Restrict2}), that is
${\tilde \gamma}=\max(0,2{\tilde \alpha}-p_2)$ on the l.h.s. and ${\tilde
 \gamma}={\tilde \alpha}$ on the r.h.s.
Thus for a given ${\tilde \alpha}$ in the range (\ref{AlphaRange})
the corresponding maximal range of ${\tilde \beta}$ is:
\begin{equation}
\label{BetaRange}
\max(0,\,2{\tilde \alpha}-p_2,\,2p_1-1,\,2p_1+p_2-1-2{\tilde \alpha})
\;\leq\; {\tilde \beta} \;\leq\; p_1
\end{equation}
In other words, if ${\tilde \alpha}$ and ${\tilde \beta}$ satisfy eqs.
(\ref{AlphaRange})-(\ref{BetaRange}), we are able
to find some~${\tilde \gamma}$ such that all probabilities
$\pi_{r_{01},r_{10},r_{11}}$ are positive.

\section{Maximal and minimal correlations}
%--------------------------------------------------
\label{AppendixMaxMin}
In order to determine the phase boundaries ${\cal B}_{max}$ and
${\cal B}_{min}$, one has to identify situations where the
correlations are extremal. For given parameters $p_1,p_2$
maximal correlations (minimal damage) simply correspond to taking the
upper bounds of the intervals (\ref{AlphaRange}) and (\ref{BetaRange}):
\begin{equation}
\tilde{\alpha}=\min(p_1,p_2)\,,
\hspace{10mm}
\tilde{\beta}=p_1\,.
\end{equation}
In case of minimal correlations (maximal damage) the situation is
more complicated since the correlation parameter $\tilde{\alpha}$
appears on the l.h.s. of eq. (\ref{BetaRange}) with a negative sign.
Therefore by increasing $\tilde{\alpha}$, the minimal value of
$\tilde{\beta}$ may decrease which makes it
impossible to predict where damage is maximal. However, we can show
that in the triangular region {\it above} 
line ${\cal L}$ in Fig. \ref{Figure2},
where $2p_1+p_2-2 \geq 0$, this problem does not arise. In this region
eq. (\ref{AlphaRange}) implies that $\tilde{\alpha} \geq p_1+p_2-1$ and
therewith $2\tilde{\alpha} \geq 2p_1+2p_2-2 \geq p_2$. Hence according to
eq. (\ref{Restrict1}) the minimal value that $\tilde{\gamma}$ can have is
$2\tilde{\alpha}-p_2$. Inserting this value into eq. (\ref{Restrict2})
we get the inequality
\begin{equation}
\max(2\tilde{\alpha}-p_2, 2p_1-1) \leq \tilde{\beta} \leq p_1
\end{equation}
which replaces eq. (\ref{BetaRange}) in the specified triangle.
In this inequality $\tilde{\alpha}$ occurs with a positive sign
on the l.h.s. and therefore the case of minimal correlations is
well defined:
\begin{equation}
\tilde{\alpha}=p_1+p_2-1\,,
\hspace{10mm}
\tilde{\beta}=2p_1-1\,,
\end{equation}
%
%
%
%
%---------------------------------------------------------------------
% Acknowledgments
%---------------------------------------------------------------------
%
%
\\[10mm]
%\noindent
{\bf Acknowledgments}\\[1mm]
H.H. would like to thank the Minerva foundation for financial support.
E.D. thanks the Newton Institute at Cambridge University and the Department
of Physics at Oxford University for their hospitality and support during
1994, when this work was started. We also thank B. Derrida
and D. Dhar for very helpful
discussions and comments, D. Stauffer for calling some useful references to 
our attention and P. Grassberger for sending us 
preprints of his work prior to publication.
%
%
%
%
%
%
%%%%%%%%%%%%%%%%%%%%%%%%%%%%%%%%%%%%%%%%%%%%%%%%%%%%%%%%%%%%%%%%%%%%%%%%%%%%
%                                REFERENCES
%%%%%%%%%%%%%%%%%%%%%%%%%%%%%%%%%%%%%%%%%%%%%%%%%%%%%%%%%%%%%%%%%%%%%%%%%%%%
%

\end{document}